***Title***: AI for Social Impact: Learning and Planning in the Data-to-Deployment Pipeline

***Authors***: Andrew Perrault, Fei Fang, Arunesh Sinha, Milind Tambe

***Abstract*****:** With the maturing of AI and multiagent systems research, we have a tremendous opportunity to direct these advances towards addressing complex societal problems. In pursuit of this goal of AI for Social Impact, we as AI researchers must go beyond improvements in computational methodology; it is important to step out in the field to demonstrate social impact. To this end, we focus on the problems of public safety and security, wildlife conservation, and public health in low-resource communities, and present research advances in multiagent systems to address one key cross-cutting challenge: how to effectively deploy our limited intervention resources in these problem domains. We present case studies from our deployments around the world as well as lessons learned that we hope are of use to researchers who are interested in AI for Social Impact. In pushing this research agenda, we believe AI can indeed play an important role in fighting social injustice and improving society.

---

The maturing of AI and multiagent systems research has created a tremendous opportunity to direct these advances towards addressing complex societal problems. In pursuit of this goal of AI for Social Impact, we as AI researchers must go beyond improvements in computational methodology; it is important to step out in the field to demonstrate social impact. To this end, we focus on three application areas: public safety and security, wildlife conservation, and public health in low-resource communities. In addition to case studies from these areas, we distill our experience into lessons learned and we hope that researchers reading this article may find them useful.

Viewing these societal problems through the lens of multiagent systems, we summarize the goals of our research program as *optimizing limited intervention resources when interacting with other agents.* The type of multiagent interaction varies widely: it might be competitive, where agents are actively trying to achieve different and often conflicting goals, or it might be a process of information spread where the agents do not have explicit goals and just passively react to their surroundings. Our overall research goal is to intervene in this multiagent interaction: to help one of the agents to achieve a desirable social objective. Towards this goal, we develop multiagent system models for the problems, such as game-theoretic models, allowing us to reason about how to maximize our limited intervention resources.

To intervene effectively, we need to understand the details of the interaction and the motivations of the different agents. However, not all elements of the interaction are known. Some elements are partially known through an often incomplete or biased dataset of observations and some are entirely unknown, requiring expert input. In the case where information gathering is time-consuming and costly, we often need to exploit available data to better understand the key latent elements and make more informed decisions.

Addressing these problems thus requires research advances in several subareas connected to multiagent systems reasoning. For example, new machine learning models are needed to analyze the data and understand the concealed aspects of the problem. Scalable optimization techniques are needed to design interventions for real-world problem instances.

We take a data-to-deployment approach to AI for Social Impact research. It begins with immersion, where we seek to understand the problem from the perspective of the decision-making agent, and ends with a field test, where we validate our modeling approach and algorithms. The data-to-deployment approach is critical because it invites us to refine our models and algorithms to enable direct social impact.

This article summarizes 12 years of work in AI for Social Impact applied to problems of public safety and security, conservation security, and public health. We provide an overview of this research: our overall research goals, the approach we have found to be successful across domains and objectives, and a history of the projects we've undertaken and their impacts.

The remainder of the article is structured as follows. We begin by defining AI for Social Impact. We then outline our solution approach: the data-to-deployment pipeline. Next, we discuss specific projects in public safety and security, conservation security, and public health and the impact these projects have had. We conclude with lessons learned and a summary.

## *Defining AI for Social Impact*

We find it useful to provide a rough definition of AI for Social Impact as a subdiscipline within AI. First, measurable societal impact should be a first-class citizen of this area of research. While a great deal of AI work can be socially beneficial, new research often has no social impact until many years later, when it is refined into a widely usable tool. In the development of computational methodologies, it is often unnecessary to think directly about the end product—expanding our knowledge and capabilities is a sufficient objective, and rightly so. In thinking about AI for Social Impact, demonstrating social impact is a key objective. Second, the research primarily focuses on vulnerable groups, e.g., disadvantaged or endangered, that lack resources to commission beneficial AI research. Third, the research area tended to have not greatly benefitted from AI research in the past. Certain problems are of great direct interest, either commercially or to governments, and as such, have been well-funded throughout the history of AI. AI for Social Impact focuses on research that would not otherwise be performed if it lacked its impact focus.

AI for Social Impact work provides value to the AI community as a whole by providing new problem models, by introducing new contexts to evaluate existing algorithms, and by raising complexities that challenge abstractions, which often motivates extensions to existing techniques. Because AI for Social Impact work requires extra effort, it requires extra considerations when evaluating its contributions. This is reflected in the AAAI 2019 and 2020 AI for Social Impact Track Call for Papers (https://aaai.org/Conferences/AAAI-20/aaai20specialtrackcall/), which states three key aspects where AI for Social Impact requires more effort than AI that focuses purely on algorithmic improvement. First, data collection may be costly and time-consuming. Second, problem modeling may require significant collaborations with domain experts. Third, evaluating social impact may require time-consuming and complex field studies. AI for Social Impact researchers invest their resources differently to make contributions to problems of great social importance.

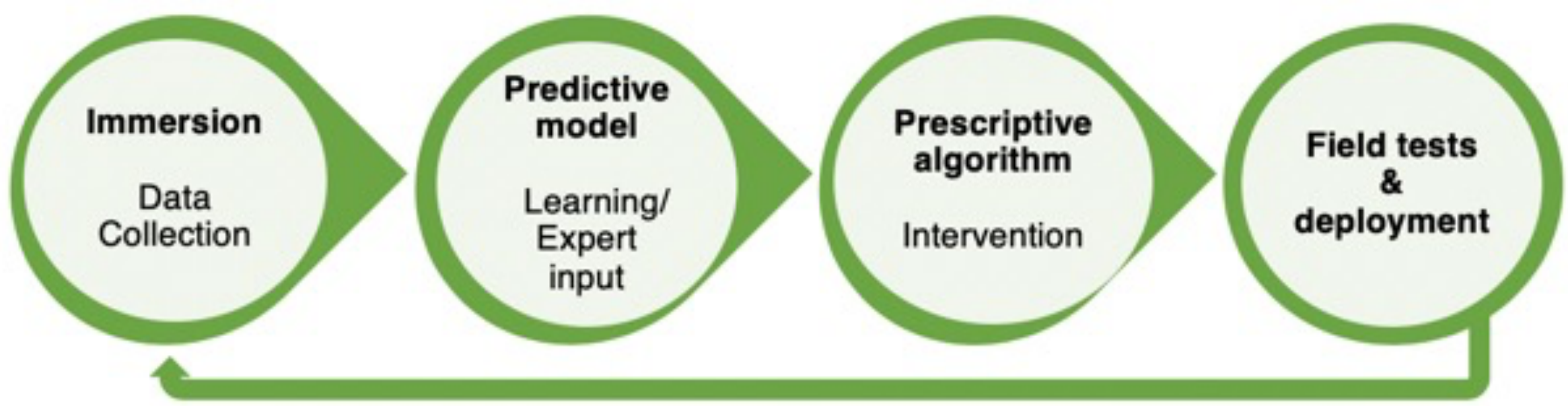

*Figure 1: The data-to-deployment pipeline that describes our approach to AI for social impact problems*

## *Solution approach: the data-to-deployment pipeline*

We characterize our solution approach as the *data-to-deployment pipeline*, which is depicted in Figure 1. Our activities at each stage of the pipeline are as follows:

### Immersion.

In the immersion stage, we seek to gather the available data about the problem and immerse ourselves in the domain. We seek to answer the following questions. First, who are the agents in the interaction? We want to understand who is making the decisions in the problem. There may be many agents, as in social network interactions, or only two, as in adversarial interactions such as basic defender-attacker interactions. Second, what information can agents use to inform their decisions? Addressing this question can be difficult for agents we do not have direct access to. We may make a pessimistic assumption when there is ambiguity: for example, in defender-attacker interactions, we assume that the adversary has access to distributional information about the defender's strategy. Third, what actions can the agents take and what impact do they have on the other agents and the environment in which they interact? What is the cost to take each action and what are the budgets? These questions may not be answerable directly but may highlight important latent aspects of the problem that may not be directly observable.[1]

We additionally gather any data that is available from past interactions: the relationships between participants, the effect of actions, the costs or rewards that were accrued, etc. During the immersion stage, we often travel to the site of the interaction and talk to the participants directly—this makes it easier to understand the perspective from the ground. We return to the interaction location in the final stage to analyze the impact of the intervention.

### Predictive model.

From the immersion stage, we understand the information flow of the interaction and what latent (unobserved) information is critical to defining the interaction. In the predictive modeling stage, we develop a strategy for handling this latent information. A common answer is to build a model that, given the data, makes predictions about high-risk vs. low-risk cases, e.g., areas that poachers may target or other classes of relevance.

**Prescriptive algorithm.**

The output of the predictive model reveals the latent state of the problem that is required to optimize our objective. In this stage, game-theoretic reasoning or multiagent systems reasoning may be used. It is often the case that an optimization problem must be solved and this may raise computational issues.

**Field tests & deployment.**

Because we take an end-to-end perspective, we must field test our solutions and compare them to the existing approach. The model we develop is necessarily a simplification of reality, and thus, field testing is the only way to confirm that we have accomplished our intended goal. This stage relates to the immersion stage, as we return to the field to evaluate our proposed solution and potentially iterate through the design process.

## *Public safety and security*

Our research program began in the domain of public safety and security. Motivated by the striking and tragic incidents of terrorist attacks in many parts of the world in the 2000s, we initiated a study of intelligent approaches to thwart attacks on public infrastructure and protect human life. We provide a brief overview of our work in this area. See Sinha et al. (2018) for a comprehensive survey.

**ARMOR: Security at LAX Airport (2007)**

Our work on patrolling the Los Angeles Airport (LAX) was described in Pita et al. (2009). We include it for completeness as it was the application that inspired this line of research. The terminals of LAX are patrolled by police to ensure the safety of passengers and the protection of infrastructure. As in most security settings, available patrollers cannot monitor every terminal simultaneously. Thus, the patrolling resources must be allocated intelligently, taking into account the differences among the terminals and the adversary's response to information gained by surveilling the patrols.

We model the problem as a *Stackelberg security game (SSG)* between the defender and an adversary (Pita et al. 2008). The defender's action is a choice from the various combinations of patrol allocations, and the adversary's action is the choice of which terminal to attack. The game's parameters, such as the value gained by the attacker and lost by the defender in the case of a successful attack, were elicited by extensive consultation with airport safety experts— these were ultimately linked to the numbers of lives potentially lost if such an attack were successful, and we were provided extensive data on passengers at different times of day in different parts of the airport.. Solving for the game's equilibrium provides the required intelligent randomized strategy. See the sidebar for a formal description of SSGs.

The deployment of our system for patrol planning at LAX, named the Assistant for Randomized Monitoring Over Routes (ARMOR), spurred extensive research activity on SSGs. As far as we know, ARMOR was the first deployed application of game theory for operational security recommendations. The successful deployment was enabled by working closely with police officers on the ground and gaining a deep understanding of the problem.

Evaluating ARMOR was especially challenging due to the (fortunate) rarity of security incidents. However, LAX police observed a significant increase in the number of firearm and drug seizures at LAX in the wake of ARMOR's deployment. While internal evaluations led the police to continue using ARMOR for the next 10 years, we provide a more thorough evaluations of deployed SSG applications through accessible data in Taylor et al. (2017).

**Stackelberg Security Game (SSG) Model and Equilibrium:**
An SSG is a game played between two players: the defender and the adversary. The defender's task is to protect $T$ targets using $k \ll T$ resources. If the adversary attacks target $i$, and $i$ is protected by the defender, the defender gets $U_d^c(i)$ and the adversary receives $U_a^c(i)$. Conversely, if not protected, the defender gets $U_d^u(i)$ and the adversary receives $U_a^u(i)$. The defender's randomized allocation (mixed strategy) results in a probability $c_i$ of covering target $i$, which results in expected utility $c_i * U_d^c(i) + (1 - c_i) * U_d^u(i)$ and $c_i * U_a^c(i) + (1 - c_i) * U_a^u(i)$ for the defender and adversary, respectively. Given a defender mixed strategy, a best responding adversary chooses a target to attack that maximizes its expected utility. Informally, the Stackelberg equilbrium is the defender mixed strategy that maximizes the defender expected utility against a best responding adversary. For a game-theoretic analysis of general Stackelberg games, see Von Stengel and Zamir (2004).

### Federal Air Marshal Service (2009).

The ARMOR application, which was featured in many news articles and was mentioned in a US Congressional subcommittee hearing, caught the attention of the Federal Air Marshal Service (FAMS). FAMS aims to deploy armed air marshals on US flights to protect passengers from dangers such as hijacking. As was the case at LAX, there are not enough marshals to cover every flight, making the problem a natural fit for modeling as an SSG. However, the defender's scheduling problem is considerably more challenging because each marshal's patrol must be a cycle. We were, once again, involved in the entire pipeline from immersion to deployment, which yielded the Intelligent Randomization in Scheduling (IRIS) system (Jain et al. 2010). IRIS was evaluated independently by the Transport Security Administration and found to be useful, and it is still in deployment today.

### PROTECT: Port and Ferry Protection Patrols (2013).

A key mission area of the US Coast Guard (USCG) is protecting ports, waterways, and coastal areas. We built the Port Resilience Operational / Tactical Enforcement to Combat Terrorism (PROTECT) system to assist the USCG to achieve this mission. One of the innovative aspects of PROTECT is ferry protection. The USCG deploys patrol boats that escort ferries, which presented new technical challenges because the ferries are mobile and the adversary's strategy space is naturally continuous. Our model was deployed to protect ferries in New York, Boston, and Houston (Shieh et al. 2012). USCG publicly released some of the data from the USCG's evaluation of PROTECT, which demonstrated that PROTECT resulted in less predictable patrolling. Furthermore, USCG reported more illicit activities within the port after PROTECT was deployed even though no additional resources were deployed.

### Rail Fare Evasion in Los Angeles.

Our work on screening rail fare evasion is an important demonstration of how the challenges of real-world deployment can motivate research. While rail fare evasion has a limited social impact, it provided an ideal testbed for evaluating the SSG approach due to a high volume of incidents and direct access to data. We began by designing a set of prescriptive patrols for transit police, as we had done in previous applications. However, when deployed, we noted that patrollers were unable to execute their assigned schedules because they were constantly being interrupted; for example, by a train running late or the need to handle a medical emergency. The feedback from deployment made us rethink our approach, leading to a sequential, Markov Decision Process-based patrolling model that accounts for execution uncertainty. The revamped model was tested on the LA subway system over 21 days in 2013 (Delle Fave et al. 2014) in an A/B test. Figure 2 summarizes the results, demonstrated that the game theoretical approach catches significantly more evaders than the status quo.

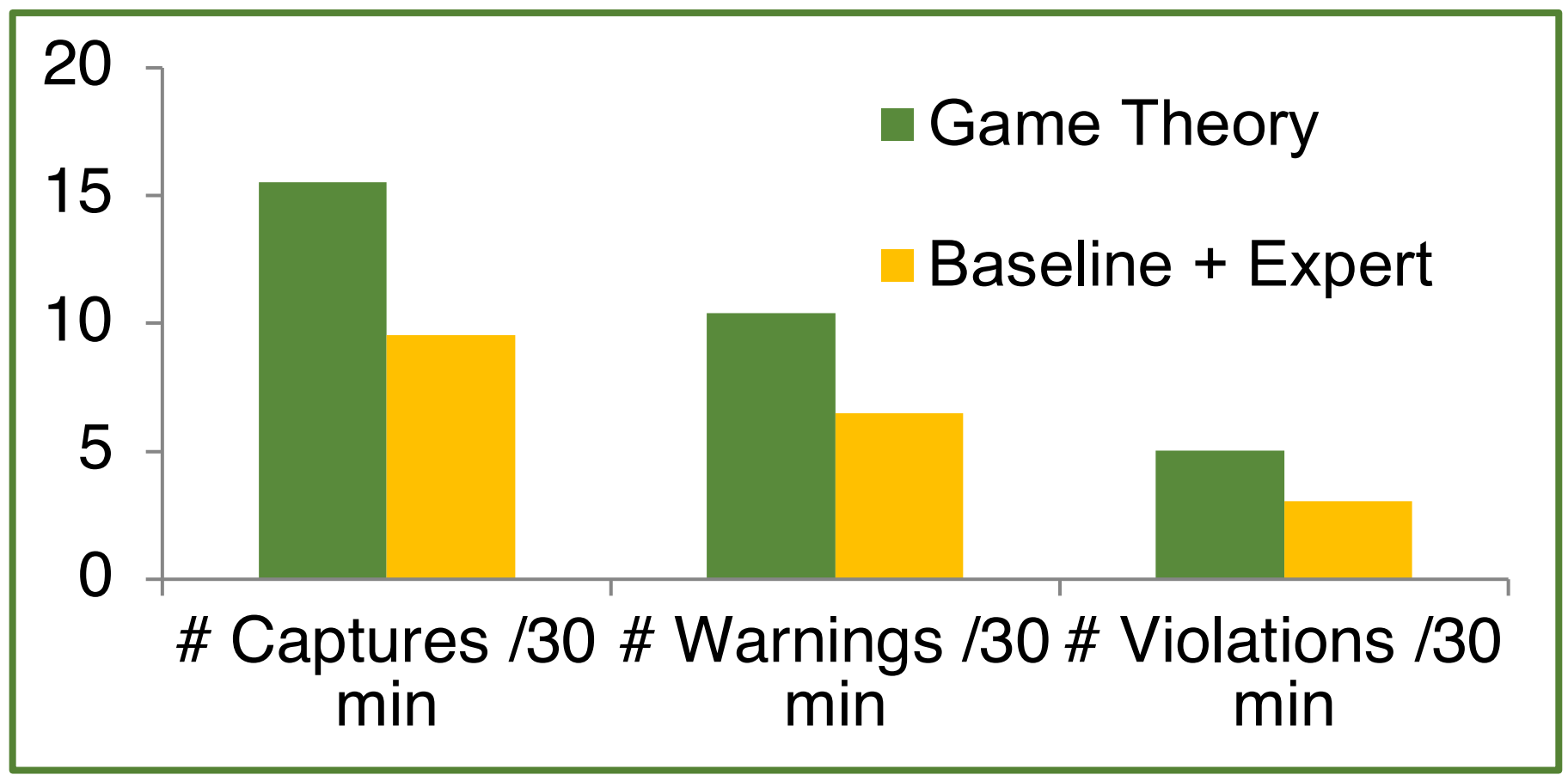

*Figure 2: Evaluation of our game-theoretic approach to rail fare evasion screening. Our model produces significantly more captures, warnings, and violations than the status quo.*

### Airport Threat Screening.

One of the more recent areas of focus in public safety and security are Threat Screening Games (TSGs), which are motivated by the problem of screening airport passengers. An adversary disguises themselves as a passenger and times their arrival to minimize the chance of detection (e.g., at a period of high screening activity and many low-risk passengers). The defender has different types of screening resources, e.g., metal detectors and advanced imaging, which screen passengers at different rates. Additionally, the defender has access to data about each passenger's risk category (the TSA constructs these based on factors such as frequency of travel) and the harm caused if the passenger were to be the adversary. The defender's goal is to balance timely screening with minimizing the chance that an adversary can slip through undetected.

Our initial formulation of TSGs imposed required that screenee must be screened in the time window they arrive in (i.e., the airport will not accept delays due to screening) (Brown et al. 2016). In this formulation, the defender's optimization is how to allocate screening resources to each category of screenees while satisfying the timing requirement. Later variations proposed more complex models: handling uncertainty in passenger arrivals and different screening rates based on the screenee. These models present the largest and hardest instance of SSGs (Xu 2016). TSGs have been tested with real-world airport data. They have also been proposed for problems outside of airport screening such as cybersecurity (Schlenker et al. 2017).

Public safety and security continue to present novel challenges as adversaries innovate. As such, defenders need to be agile, making use of AI tools to reflect the realities of a changing threat environment.

## *Conservation security*

The successes in public safety and infrastructure security inspired us to consider what we call conservation security domains which also feature limited law enforcement resources. Illegal activities such as poaching, illegal logging, and illegal, unreported, and unregulated fishing can lead to the destruction of ecosystems. For example, the African elephant population declined by 30 percent between 2007 and 2014 primarily due to illegal poaching. To combat such activities, law enforcement sends patrollers as well as more advanced tools, such as aircraft and drones, to areas of interest to detect and deter illegal activities. However, the patrolling resources are even sparser than those in the public safety and security domain. For example, at one point, only 60 rangers were patrolling Murchison Falls National Park (MFNP) in Uganda, which is almost 4000 square kilometers.

The role of data is dramatically different in conservation security than in counter-terrorism tasks mentioned earlier. First, there is much more data available. For example, rangers at MFNP remove more than a thousand snares per year (**Error! Reference source not found.**). They record their patrol routes and the locations of snares using the Spatial Monitoring and Reporting Tool (SMART), creating data that can be analyzed. Second, the data is uncertain in multiple ways—for example, rangers may fail to find a snare even if one is present. The central role of data makes the interaction between game theory and machine learning a key aspect of conservation security research. In this section, we describe PAWS and SPOT: two conservation security projects that have traversed the data-to-deployment pipeline.

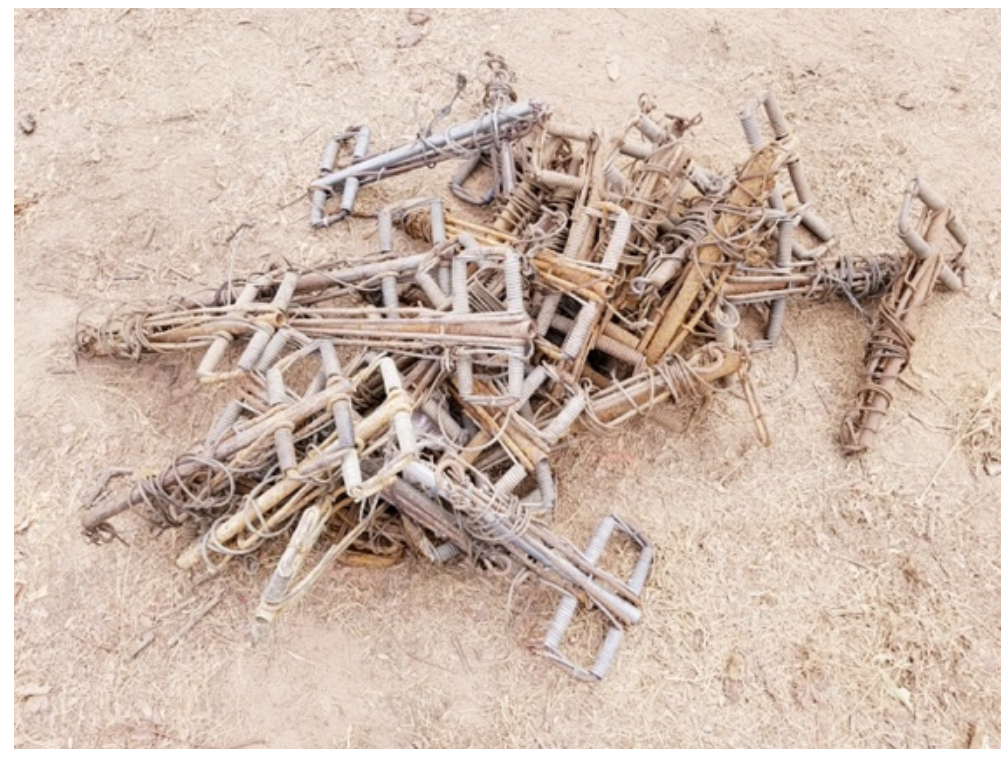

*Figure 3: Snares found in Srepok Wildlife Sanctuary, Cambodia, the location of the latest field test of PAWS.*

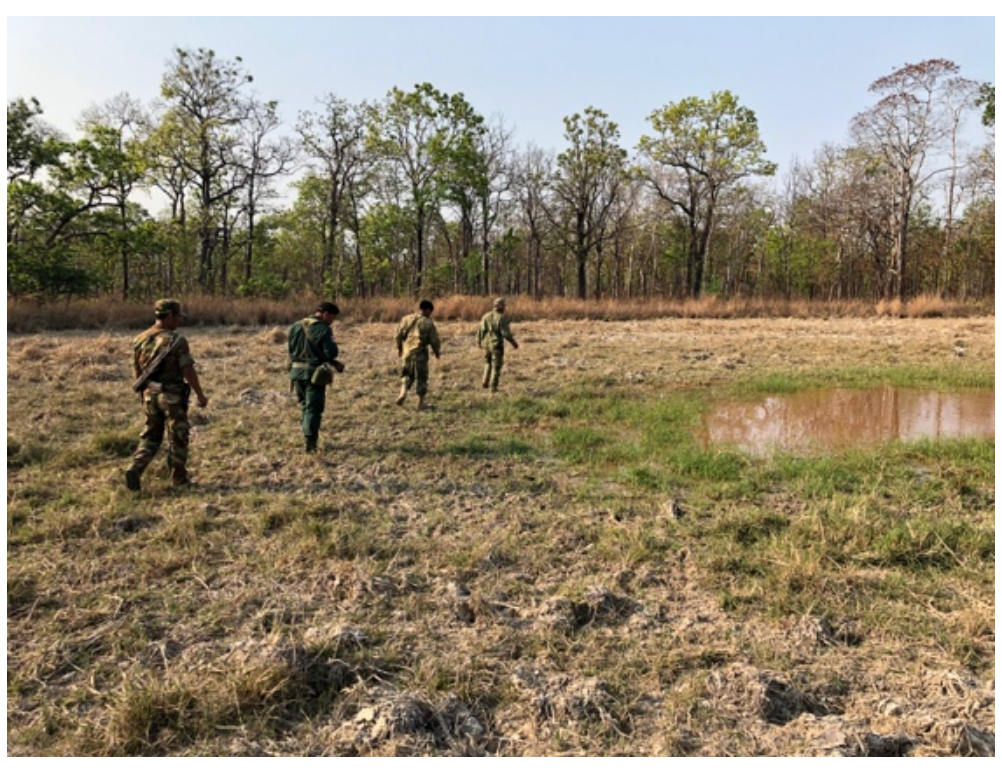

*Figure 4: Rangers on patrol in Srepok Wildlife Sanctuary, Cambodia.*

## PAWS: The Protection Assistant for Wildlife Security.

The Protection Assistant for Wildlife Security (PAWS) is our system for predicting poaching threats and planning ranger patrols to combat poaching. The system consists of three modules: (i) a model to predict poaching behavior; (ii) a game-theoretic model for coarse-grained patrol optimization; (iii) a fine-grained patrol planner that takes into account detailed terrain information. Each module has gone through several iterations, and we elaborate on the key developments. PAWS is now being integrated into the SMART tool, which has been adopted by more than 800 protected areas worldwide.

In module (i), we aim to leverage the available data to predict the intensities of poaching activities. Initial versions of this model extended the behavioral game-theoretic approach developed in the public safety setting (Fang et al. 2014), calculating the subjective utility of poachers as a linear combination of feature values of each target. A target is a cell in a 1 km by 1 km grid representing the protected area. The features of a target may include historical and current patrol effort as well as geospatial features such as animal density, land cover, and slope. A label indicates whether or not poaching activity was found in the corresponding cell at a particular time.

This approach was only partially successful when applied to real-world data in Queen Elizabeth National Park (QENP). First, there were very few positive examples relative to the size of the park. Second, we did not handle uncertainty in the data since a ranger may fail to find a snare even if one is present. More recent work uses more sophisticated machine learning techniques to address these challenges. For example, Gholami et al. (2018) trains a different classifier for each level of patrol effort and combines them in an ensemble, achieving better predictive accuracy as a result.

We performed extensive validation of the learned models. Our first test sent rangers to two areas in QENP predicted to be poaching hotspots but that were not frequently patrolled (Kar et al. 2016). The rangers found three sets of snares in a month, outperforming 91% of historical months. Following that success, we conducted an 8-month field test where rangers were sent to 27 areas predicted to be either high or low threat by our model. We found that the catch per unit effort, i.e., the number of snares found per km of walking, was ten times higher in the regions that were predicted to be high threat than those predicted to be low threat. Later experiments in different protected areas confirmed that our model is effective at identifying and predicting poaching hotspots.

In module (ii), we build a game-theoretic model of the interaction between the rangers and the poachers and use it to design patrol strategies that maximize the defender's utility (Xu et al. 2017). We treat the learned model from module (i) as a black box that describes the adversary's behavior, taking the proposed patrol effort and target features as inputs and yielding the probability that a snare will be discovered. The resulting optimization problem is to maximize the expected number of snares discovered by the defender subject to the defender's scheduling constraints, namely that the patroller always starts from the patrol post and must return to it at the end of the patrol and that patrols have limited distance. We solve this model using mixed-integer linear programming.

While module (ii) considers coarse scheduling constraints, the actual patrols often need to satisfy more fine-grained constraints—complex terrain may make it impossible for rangers to move from one grid cell to another. In module (iii), we incorporate terrain information by building a virtual street map of the area and constructing the patrol strategy on this map (Fang et al. 2016). This module was key to the success in a field test in Malaysia where multiple signs of human and animal activity were found.

An avenue for future improvement of PAWS is to consider the interaction between the prediction and game-theoretic models. Our recent work in game-focused learning (Perrault et al. 2020) has shown that including a game model in the machine learning pipeline improves the defender's utility.

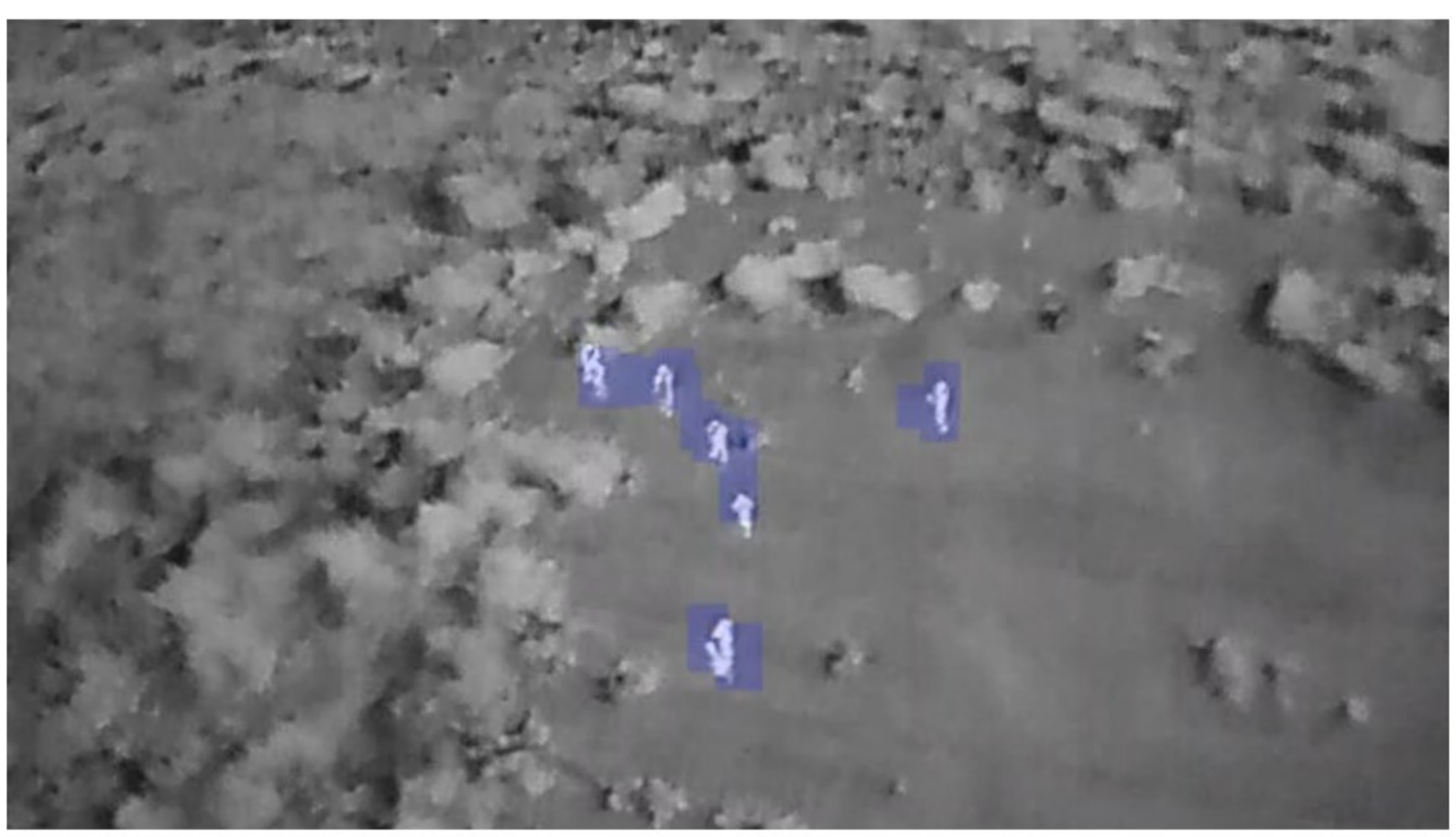

*Figure 5: SPOT was able to detect humans in a test run by NGO partner AirShepherd.*

**SPOT: Systematic POacher deTector for conservation drones.**

Drones can be a valuable patrolling tool. They can be equipped with long-wave thermal infrared cameras, allowing them to effectively detect poachers at night when many poachers are active. The video is then transmitted in real time to ranger stations. Drones present three main technical challenges. First, monitoring drone-captured video is tedious and challenging. Second, drones cannot directly interdict the poachers and force them to leave the area. Therefore, the drones and rangers must be coordinated. Third, drones can display a flashing light, alerting poachers that they are being observed. This signaling capability, if used carefully, can dissuade poaching activity through the threat that a ranger will be dispatched. However, if overused, signals lose credibility and poachers ignore them.

SPOT (the Systematic POacher deTector) is designed to tackle the first challenge. It augments conservation drones with the ability to automatically detect humans and animals in near real time (Bondi et al. 2018). Given historical videos taken by UAVs, we treat each video frame as an image and collect labels (bounding boxes) for any humans or animals. Our deep learning-based model leverages available computing resources (e.g., GPU laptops, cloud computing) to improve the detection speed of SPOT in the field. Air Shepherd, a drone-based conservation program, conducted a real-world test, with promising results.

To plan the coordination of drones and human patrollers as well as the signaling scheme, we build a Sensor-Empowered Security Game (SEG) model based on SSGs (Xu et al. 2018). We show that, in the optimal signaling scheme, the drones always send a warning signal when there is a nearby ranger and send a deceptive warning signal with a carefully designed probability when there is no nearby patroller. Simulation results show that well-coordinated deployment and signaling significantly benefits the rangers. This model assumes that drones always detect a poacher when one is present, and we are currently working to extend the model to account for detection uncertainty.

## *Public health*

In this section, we describe two major public health projects we have undertaken. The first focuses on spreading information to prevent HIV among homeless youth in Los Angeles. The second aims to improve tuberculosis medication adherence in India.

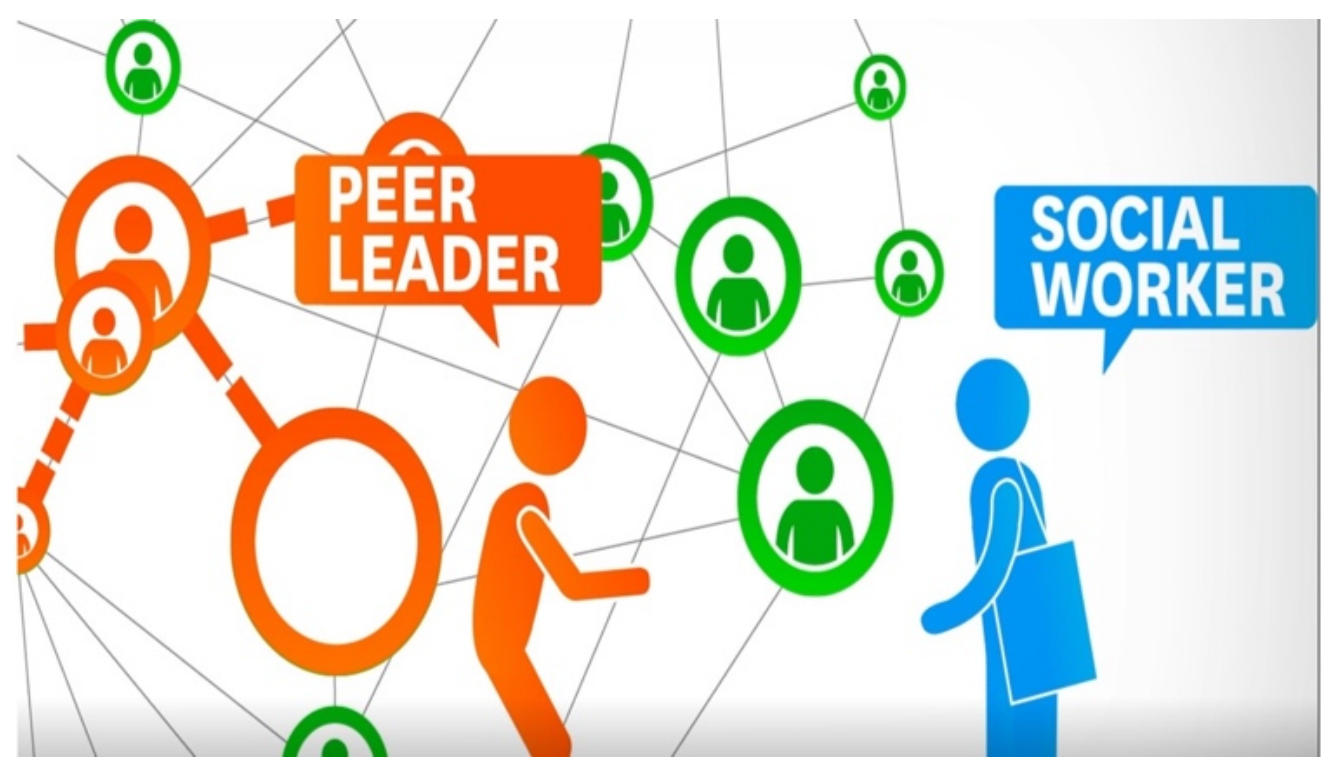

*Figure 6: Social workers educate peer leaders about HIV prevention, information which the peer leader is to disseminate in their social network.*

### Preventing the spread of HIV among homeless youth.

Homelessness affects around 2 million youths in the United States annually, 11 percent of whom are infected with the human immunodeficiency virus (HIV), which is 10 times the rate of infection in the general population (Aidala and Sumartojo 2007). Peer-led HIV prevention programs such as Popular Opinion Leader (POL) (Kelly et al. 1997) try to spread information about HIV prevention through a social network of homeless youth by identifying peer leaders within the network to champion the message. The traditional strategy for selecting peer leaders is via “degree centrality”—that is, nodes with the highest number of friendships are picked first. Such peer-led programs are highly desirable to agencies working with homeless youth as these youth are often disengaged from traditional health-care settings and are distrustful of adults. Strategically choosing intervention participants is important so that information percolates through their social network in the most efficient way.

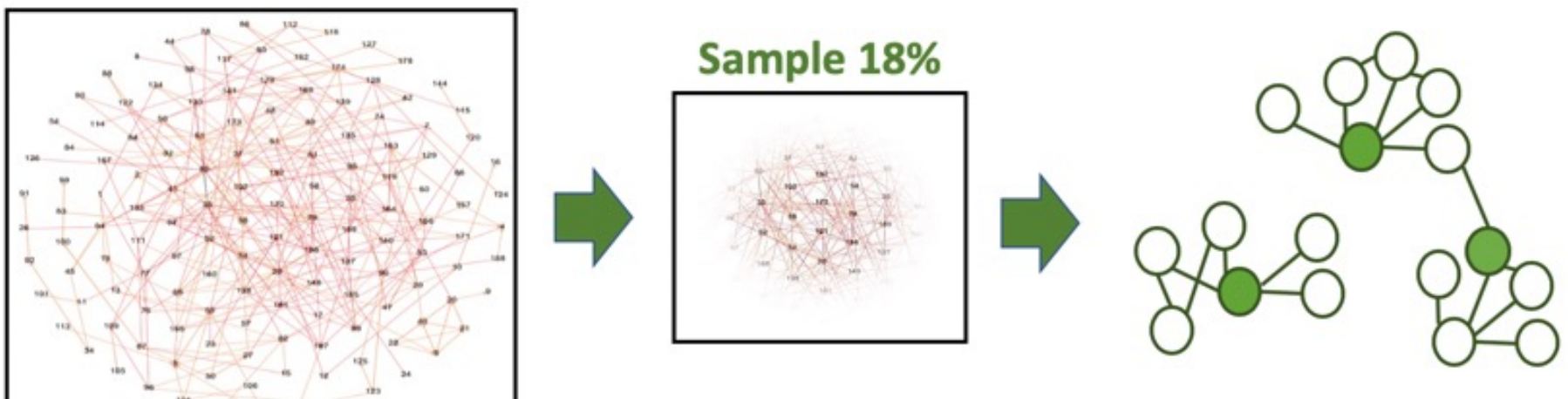

*Figure 7: We decide how to spread HIV prevention information across a network by sampling a small number of edges.*

We formulate the problem of selecting peer leaders to spread HIV prevention information as influence maximization with uncertain parameters over an uncertain network. We assume that the underlying process that is spreading information is an independent cascade model (Kimura and Saito 2006) on a graph $G = (V, E)$ and an associated function $f(v)$, which represents the probability that influence spreads across edge $v$. We are uncertain about $f(v)$ and want to maximize the number of influenced nodes in a robust way. We show that we can achieve this objective by formulating the problem as a game against nature, where nature chooses $f$ in response to our choice of seeds, then solving it via double oracle (Wilder et al. 2017). This approach yields an equilibrium strategy despite the exponential search space for the players and converges with approximation guarantees.

A further complication that arises in practice is the unavailability of peer leaders that we select. For instance, a youth may have gotten arrested or gone to stay with relatives. Thus, we choose to think about the problem as choosing a set of peer leaders each week for many weeks according to a training budget. In each successive week, we discover which youth were able to participate last time, informing which new youths to invite this week to continue to maximize information spread. The resulting problem can be formulated as a POMDP and solved via POMDP decomposition, yielding the HEALER algorithm (Yadav et al. 2015).

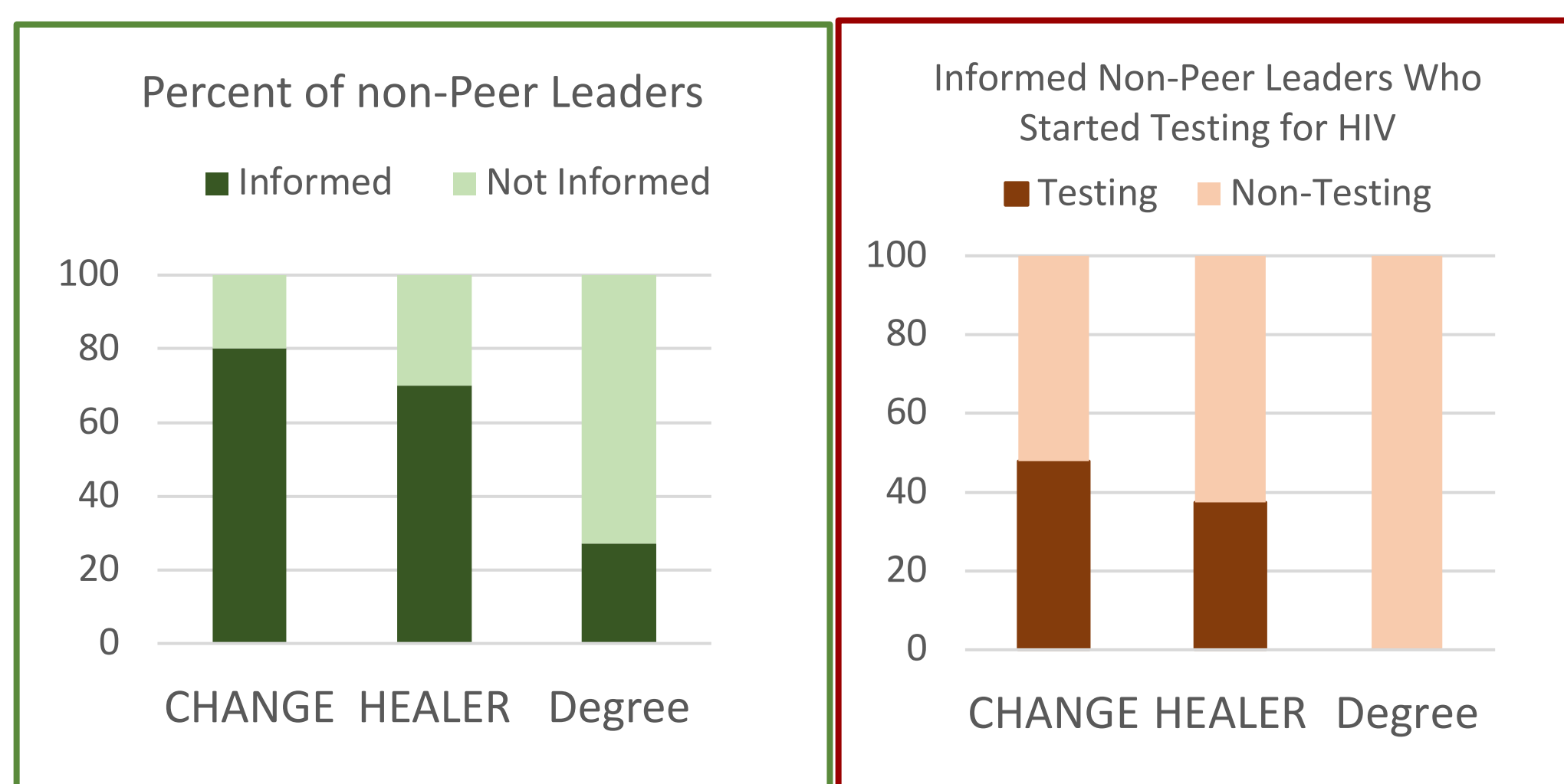

*Figure 8: These charts compare three approaches to selecting peer leaders to spread information about HIV prevention. HEALER, our algorithm that uses network structure to select nodes, outperforms degree centrality (Degree) in both the percent of non-peer leaders reached and the percent of non-peer leaders who began testing for HIV. CHANGE, which uses only partial network information, performs as well as HEALER at a lower surveying cost.*

We performed a pilot field test of HEALER, comparing it to the most popular baseline of degree centrality. We selected communities of 60 youths at different centers for homeless youth and our collaborators in social work trained 12 of those youths to be peer leaders (Rice et al. 2018). HEALER is significantly more effective at spreading information in these tests—it reaches around 75% of non-peer leaders, compared to only 25% for degree centrality (see Figure 8). As a result, HEALER is more effective at causing youth to start testing for HIV: around 30-40% of the community began testing, compared to 0% for degree centrality.

However, despite its greater effectiveness, HEALER incurs higher costs than degree centrality because it requires that the entire social network be surveyed via on-the-ground work by social workers over many weeks. To overcome this obstacle, we develop a variant of HEALER that only surveys the connections among a small subset of youth (Wilder et al. 2018a). This algorithm, CHANGE, performed as well in field tests as HEALER (see Figure 8), while surveying only 18% of the youth in the network—a major cost reduction.

In other work, we have modeled social influence over a network to optimize public health objectives including preventing childhood obesity in the Antelope Valley (Wilder et al. 2018b) and preventing suicide among college students (Rahmattalabi et al. 2019b).

### Ensuring tuberculosis medication adherence.

Tuberculosis (TB) is one of the top ten causes of death worldwide and is the deadliest infectious disease*;* last year alone approximately 10 million people across the globe were infected with TB, leading to 1.8 million deaths. The prevalence of TB is partly attributable to its disproportionate effect on the world's global south where the poor have extremely limited access to healthcare, clean living conditions, and education, which all contribute to the spread of the disease. Further, multi-drug resistant strains of TB, which are far more expensive and difficult to treat than drug-susceptible TB strains, have taken hold in the world's global south.

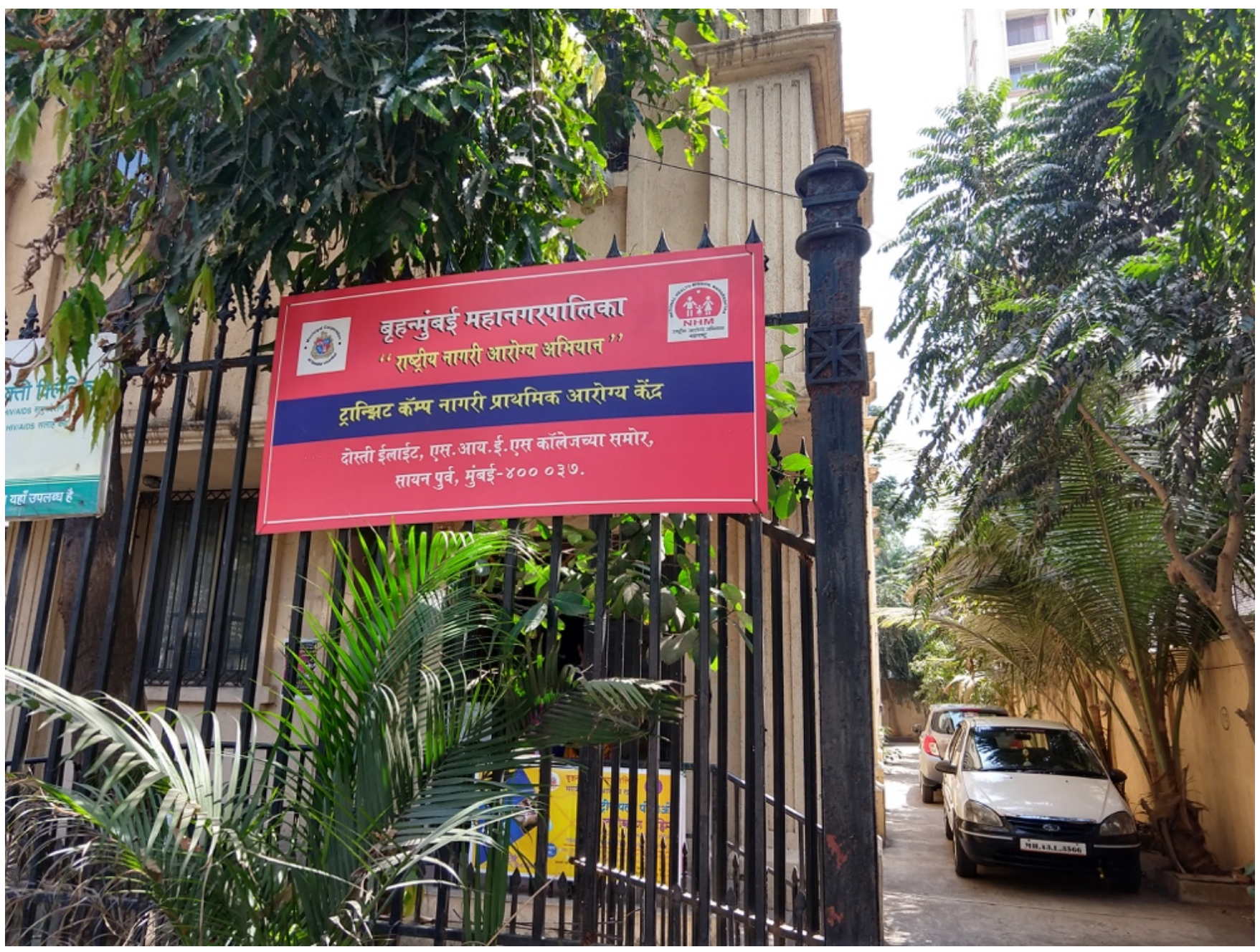

*Figure 9: TB control center in Mumbai, India.*

The prevalence of TB is caused in part by non-adherence to medication, resulting in a greater risk of death, reinfection, and contraction of drug-resistant TB. To combat non-adherence, the WHO recommends directly observed treatment (DOT), in which a health worker confirms that a patient is consuming the required medication daily by observing the patient taking the medication. However, requiring patients to travel to the DOT facility imposes a financial burden and potential social stigma due to public fear of the disease. Such barriers contribute to patients dropping from treatment, making TB eradication difficult. Thus, digital adherence technologies (DATs), which give patients flexible means to prove adherence, have gained global popularity (Subbaraman 2018).

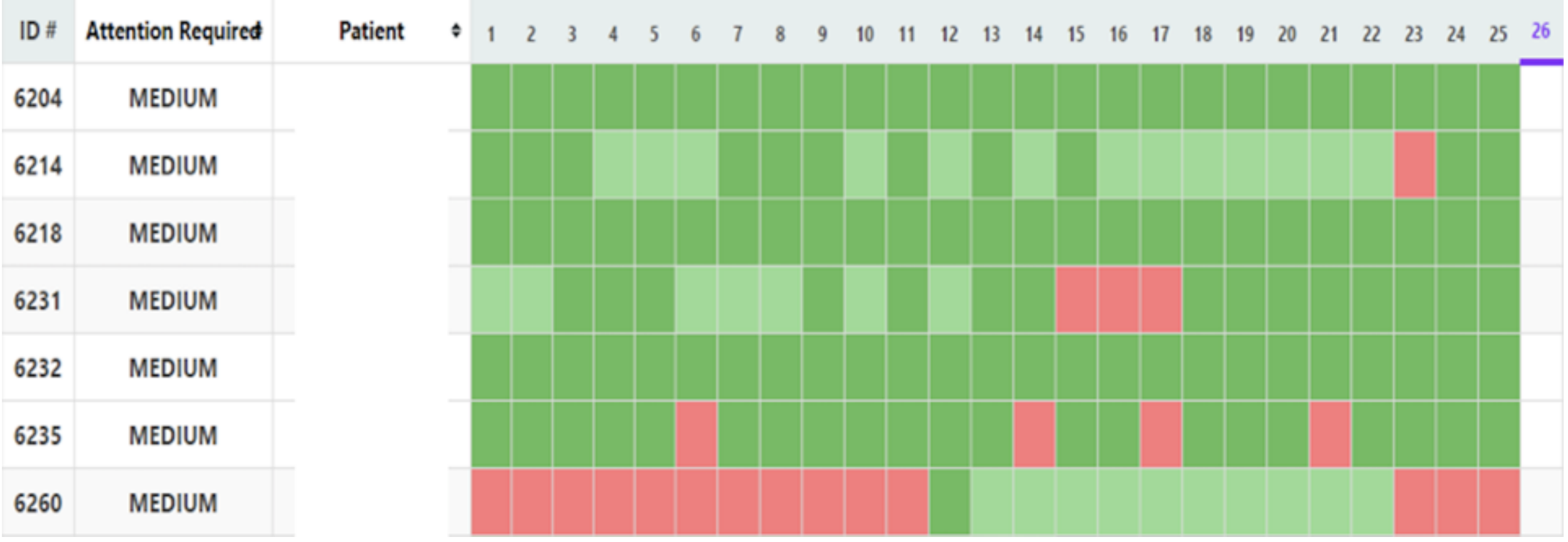

*Figure 10: 99DOTS electronic adherence dashboard seen by health workers for a given month. Missed doses are marked in red while consumed doses are marked in green.*

DATs allow patients to be "observed" consuming their medication electronically, e.g. via two-way text messaging, video capture, electronic pillboxes, or toll-free phone calls. Health workers can then view real-time patient adherence on a dashboard such as the one seen in Figure 10. In addition to improving patient flexibility and privacy, the dashboard enables health workers to triage patients and focus their limited resources on the highest risk patients.

Our objective is to use the longitudinal data collected by DATs to help health workers better triage TB patients and deliver interventions to boost the overall adherence of their cohorts (Killian et al. 2019). At first glance, the problem of predicting whom to target for an intervention appears to be a simple supervised machine learning problem. Given data about a patient's medication adherence, one can train a machine learning model to predict whether they will miss medication doses in the future. However, such a model ignores the concurrent interventions from health workers as the data were collected and can lead to incorrect prioritization decisions even when it is highly accurate. For instance, we might observe that missed doses are followed by a period of medication adherence: this does not mean that people with missed doses are more likely to take medication but, most likely, that there was an intervention by a health worker after which the patient restarted their medication.

We introduce a general approach for learning from adherence data with unobserved interventions, based on domain knowledge of the intervention rules applied by health workers. Using data from the DAT operated by the City TB Office of Mumbai, we show that our approach enables health workers to identify 21% more high-risk patients and catch 76% more missed doses than the currently used heuristics.

We can further improve outcomes by using an end-to-end, decision-focused learning approach (Wilder et al. 2019a). Such approaches focus on making predictions that induce good downstream decisions—such as choosing patients for interventions—rather than making perfectly accurate predictions about adherence. In our setup, this approach tunes our system to be more accurate among those patients who could benefit from intervention, rather than being equally accurate across all patients. We find that such a classifier improves the number of successful interventions by approximately 15% compared to a non-decision-focused approach, despite being less accurate about future medication adherence.

## *Lessons learned*

Based on the experience of the work discussed so far, we state six broad lessons that we have found generally useful. The first two are philosophical (what perspective should we take as AI for Social Impact researchers), the third is technical, and the remainder relate to the multidisciplinary nature of AI for Social Impact work.

### Take a data-to-deployment perspective.

We select projects that can lead directly to real-world deployment in the near future. An academic approach that emphasizes improvements in computational methodology is not necessarily well-suited to achieving this goal—we need to be able to take all the steps from accessing relevant data to deploying prototypes in the field.

### Go out into the field.

Often AI for Social Impact entails working with vulnerable communities and in remote areas. It is difficult to understand the problems we are trying to solve without consulting the users in the field directly and eliciting crucial details that would not have come to light in the laboratory setting. Additionally, visiting a site allows researchers to understand what technological resources (e.g., level of computing power, connectivity) will be available to the intended end-user of the AI solution.

### Lack of data is the norm and needs to be addressed in the project strategy.

It is rarely the case that sufficient data exists in a social impact setting and developing strategies to address the lack of data is a critical element of our work. For an example project where we apply these strategies, see our project on preventing the spread of HIV among homeless youth.

The first strategy is to make data acquisition part of the deployment plan. If a partner is sufficiently motivated to implement an AI solution, collecting data can energize people working on the ground. Collecting data about the existing interaction between agents on the ground is the first step in adapting to an AI approach.

The second strategy is to make data acquisition part of the technical contribution of the project. If data is difficult to acquire, choosing how to collect it can be part of the AI problem (e.g., through active learning, preference elicitation or reinforcement learning). For a solution to be sustainable, the cost of collecting the necessary data must be less than the benefit the solution provides.

The third strategy is to consider sparse data when selecting algorithms. For example, much recent progress in machine learning has focused on cases where there is a large amount of labeled or unlabeled data available. When these conditions are not met, older, statistical approaches may perform better.

The fourth strategy is to consider expert input or human subject experiments. In some circumstances, data is so rare, expensive or sensitive that techniques driven by real-world data are not suitable. This problem arises especially in public security settings where attacks can rarely be observed.

### AI for Social Impact work should be evaluated differently than other AI areas.

Significant amounts of time and effort must be spent on developing partnerships, modeling, and evaluation to perform research that has a concrete near-term impact. These areas of emphasis require a different approach to evaluation compared to the one traditionally used at AI conferences.

### Build interdisciplinary partnerships.

AI for Social Impact work cannot be done without partnerships with researchers in other disciplines who are experts on social impact problems. AI researchers are, by necessity, primarily focused on the problems that arise from the perspective of AI methodology. Thus, if AI is to have a real-world positive impact, it is necessary to leverage expert perspectives on the problems we are trying to address.

### Fairness: an emerging concern.

In research done so far, fairness has been a part of the ethos of partner organizations. As they have been more aware of the challenge of bias in AI systems, questions of fairness have been arising in our research. Some of these questions are quite complicated. While we are currently exploring algorithmic solutions to some of the questions raised (Wilder et al. 2019b, Rahmattalabi et al. 2019a), a key question for future investigation is to understand the interaction between domain-specific stakeholder perspectives on fairness and algorithmic approaches.

## *Summary*

Looking to the future, we believe AI is important for improving society and fighting social injustice. To that end, in pushing forward the agenda of AI for Social Impact, we need to engage in interdisciplinary collaborations and bring the benefits of AI to populations that have not benefited from it. We hope that the case studies we provided and the insights we have gathered are useful.

1. In many other disciplines, such as human-computer interaction and social work, descriptive work is publishable on its own (e.g., Ismail and Kumar, 2019) and may be used as a jumping off point for intervention design (Fraser and Galinski, 2010). In AI, the descriptive work performed in the immersion stage is a necessary prerequisite for building an AI system, but would not generally be publishable in an AI venue unless paired with the deployment of an intervention.

***Bios***

**Andrew Perrault** is a postdoctoral fellow at the Center for Research on Computation and Society at the John A. Paulson School of Engineering and Applied Science at Harvard University. His research is in the field of artificial intelligence and multiagent systems, focusing on multiagent interactions where information about the agents is not known a priori and must be queried or learned from data.

**Fei Fang** is an Assistant Professor at the Institute for Software Research in the School of Computer Science at Carnegie Mellon University. Her research lies in the field of artificial intelligence and multi-agent systems, focusing on integrating learning with game theory and contributing to the theme of AI for Social Good. Her work has won multiple awards, including Distinguished Paper at IJCAI-ECAI'18, Innovative Application Award at IAAI'16, the Outstanding Paper Award in Computational Sustainability Track at IJCAI'15.

**Arunesh Sinha** is an Assistant Professor in the School of Information Systems at Singapore Management University. His research is at the intersection of security, machine learning, and game theory. His interests lie in the theoretical aspects of multi-agent interaction, machine learning, security, and privacy, along with an emphasis on the real-world applicability of the theoretical models. His paper was nominated for the best application paper at AAMAS 2016. He was awarded the Bertucci fellowship in his Ph.D. at Carnegie Mellon University in appreciation of his novel research.

**Milind Tambe** is Gordon McKay Professor of Computer Science and Director of Center for Research on Computation and Society at Harvard University; concurrently, he is also Director "AI for Social Good" at Google Research India. Prof. Tambe's research focuses on advancing multiagent systems research for Social Good. He is the recipient of the IJCAI John McCarthy Award, ACM/SIGAI Autonomous Agents Research Award and AAAI Robert S. Engelmore Memorial Lecture award. Prof. Tambe is a fellow of ACM and AAAI. The work reported in this article was performed prior to Prof. Tambe's joining Google.